\documentclass[aps,preprint,groupedaddress,showpacs,floatfix]{revtex4}
\usepackage{graphicx}
\begin{document}

\title{\boldmath $X(3872)$, $I^G(J^{PC})=0^+(1^{++})$, as the
$\chi_{c1}(2P)$ charmonium }
\author {
N.N. Achasov$^{\,a}$ \email{achasov@math.nsc.ru} and E.V.
Rogozina$^{\,a,b}$ \email{rogozina@math.nsc.ru}}

\affiliation{
   $^a$Laboratory of Theoretical Physics,
 Sobolev Institute for Mathematics, 630090, Novosibirsk, Russia\\
$^b$Novosibirsk State University, 630090, Novosibirsk, Russia}

\date{\today}

\begin{abstract}

 Contrary to almost standard opinion that the $X(3872)$ resonance
is the $D^{*0}\bar D^0+c.c.$ molecule or the $qc\bar q\bar c$
four-quark state, we discuss the scenario where the $X(3872)$
resonance is the
 $c\bar c = \chi_{c1}(2P)$ charmonium which "sits on" the  $D^{*0}\bar D^0$ threshold.

We explain the shift of  the mass of the
 $X(3872)$ resonance with respect to the prediction of a potential model for the mass of the
$\chi_{c1}(2P)$ charmonium by the contribution of the virtual
$D^*\bar D+c.c.$ intermediate states into the self energy of the
$X(3872)$ resonance.  This allows us to estimate the coupling
constant of the $X(7872)$ resonance with the $D^{*0}\bar D^0$
channel, the branching ratio of the $X(3872) \to D^{*0}\bar D^0 +
c.c.$ decay, and the branching ratio of  the $X(3872)$ decay into
all non-$D^{*0}\bar D^0 + c.c.$ states. We predict a significant
number of unknown decays of    $X(3872)$ via two gluon:
$X(3872)\to gluon\ gluon\to hadrons$.

 We suggest a physically clear program of experimental researches for
verification of our assumption.

\end{abstract}
\maketitle

The $X(3872)$ resonance became the first in discovery of the
resonant structures $XYZ$ ($X(3872)$, $Y(4260)$, $Z_b^+(10610)$,
$Z_b^+(10650)$, $Z_c^+(3900)$), the interpretations of which as
hadron states assumes existence in them at least pair of heavy and
pair of light quarks in this or that form. Thousand articles on
this subject already were published in spite of the fact that many
properties of new resonant structures are not defined yet and not
all possible mechanisms of dynamic generation of these structures
are studied, in particular, the role of  the anomalous Landau
thresholds is not studied. Anyway, this spectroscopy took the
central place in physics of hadrons.

Below we give reasons that  $X(3872)$, $I^G(J^{PC})=0^+(1^{++})$,
is the $\chi_{c1}(2P)$ charmonium and suggest a physically clear
program of experimental researches for verification of our
assumption.

 The two dramatic discoveries have generated a stream of the
$D^{*0}\bar D^0+D^0\bar D^{*0}$  molecular interpretations of the
$X(3872)$ resonance.

The mass of the $X(3872)$ resonance is 50 MeV lower than
predictions of the most lucky naive  potential models for the mass
of the $\chi_{c1}(2P)$ resonance,
\begin{equation}
\label{shiftmass} m_X-m_{\chi_{c1}(2P)}= -\Delta\approx -
50\,\mbox{MeV},
\end{equation}
and the relation between the branching ratios
\begin{equation}
\label{isotopicviolation}
 BR(X\to\pi^+\pi^-\pi^0J/\psi(1S))\sim
BR(X\to\pi^+\pi^-J/\psi(1S))\,,
\end{equation}
 that is interpreted as a strong
violation of isotopic symmetry.

  But the bounding energy is small,
$\epsilon_B\lesssim (1 \div 3)$ MeV. That is, the radius of the
molecule is large, $r_{X(3872)}\gtrsim (3 \div 5)$ fm $ = (3 \div
5)\cdot10^{-13}$ cm. As for the charmonium, its radius is less one
fermi, $r_{\chi_{c1}(2P)}\approx 0.5$ fm $=0.5\cdot 10^{-13}$ cm.
 That is, the molecule volume is $100 \div 1000$ times as large
as the charmonium volume, $V_{X(3872)}/V_{\chi_{c1}(2P)}\gtrsim
100 \div 1000$.

How to explain sufficiently abundant inclusive production of the
rather extended molecule X(3872) in a hard process $pp\to X(3872)
+ anything$ with rapidity in the range 2,5 - 4,5 and transverse
momentum in the range 5-20 GeV \cite{LHCb12}? Really,
\begin{equation}
\label{pptoX(3872)}
 \sigma (pp \to X(3872) + anything)
BR(X(3872)\to\pi^+\pi^-J/\psi)=5.4\, \mbox{nb}
\end{equation}
 and
\begin{equation}
\label{pptopsi(2S)} \sigma (pp\to\psi(2S) + anything)
BR(\psi(2S)\to\pi^+\pi^-J/\psi) =38\, \mbox{nb}.
\end{equation}
But, according to Ref. \cite{PDG14}
\begin{equation}
\label{psi(2S)Jpsi}
 BR(\psi(2S)\to\pi^+\pi^-J/\psi)=0.34
\end{equation}
   while
\begin{equation}
\label{X(3872)Jpsi}
 0.023<BR(X(3872)\to\pi^+\pi^-J/\psi)<0.066
\end{equation}
according to Ref. \cite{Belle09}.
 So,
\begin{equation}
\label{pptoX(3872vspptopsi(2S)}
 0.74<\frac{\sigma (pp \to X(3872)
+ anything)}{ \sigma (pp\to\psi(2S) + anything)}<2.1.
\end{equation}
 The
extended molecule is produced in the hard process as intensively
as the compact charmonium. It's a miracle.

\begin{figure}[h]
\begin{center}
\includegraphics[width=10cm,height=6cm]{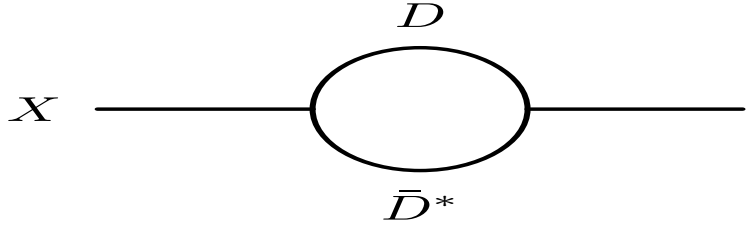}
\end{center}
\vspace*{-45pt} \caption{The contribution of the $\bar D^0 D^{*0}$
and  $D^- D^{*+}$ loops into the self energy of the $X(3872)$
resonance.} \label{fig1}
\end{figure}

As for the problem of the mass shift, Eq. (\ref{shiftmass}), the
contribution of the $D^- D^{*+}$ and $\bar D^0 D^{*0}$ loops, see
Fig. \ref{fig1}, into the self energy of the $X(3872)$ resonance,
$\Pi_X(s)$, solves it easily.

\begin{equation}
  \Pi_X(s)=\Pi_X^{\bar D^0
D^{*0}}(s)+ \Pi_X^{D^- D^{*+}}(s)=\frac{g_A^2}{8\pi^2}\left
(I^{\bar D^0 D^{*0}}(s)+I^{D^- D^{*+}}(s)\right )\,,
\end{equation}
where
\begin{equation}
   I^{D\bar
D^*}(s)=\int\limits_{m_+^2}^{\Lambda^2}
\frac{\sqrt{(s'-m_+^2)(s'-m_-^2)}}{s'(s'-s)}ds'\approx 2
\ln\frac{2\Lambda}{m_+}
-2\sqrt\frac{m_+^2-s}{s}\arctan\sqrt{\frac{s}{m_+^2-s}}\,,
\end{equation}
where
\begin{equation}
m_+=m_{D^*}+m_D\,,\ \ m_-=m_{D^*}-m_D\,,\ \ s<m_+^2\,,\ \
\Lambda^2\gg m_+^2.
 \end{equation}

  For the calculations we use the Lagrangian
\begin{eqnarray}
\label{L}
 && L(x)=g_AX^\mu\left(D_\mu(x)\bar D(x)+ \bar
D_\mu(x)D(x)\right)\nonumber\\[6pt]
 &&
=g_AX^\mu\Bigl(D^0_\mu(x)\bar D^0(x)+ \bar D^0_\mu(x)D^0(x) +
D^+_\mu(x)D^-(x)+ D^+_\mu(x)D^-(x)\Bigr ).
\end{eqnarray}
 The width of the $X\to D^{*0}\bar D^0 + c.c.$
decay
\begin{equation}
\label{width}
 \Gamma(X \to D^{*0}\bar D^0 + c.c.\,,\, s )\approx
(g_A^2/8\pi)(2|\vec{k}|/s)\,.
\end{equation}
 The inverse propagator of the X(3872) resonance
\begin{equation}
\label{propagator}
 D_X(s)= m_{\chi_{c1}(2P)}^2-s -\Pi_X(s)-\imath
m_X\Gamma\,,
\end{equation}
where $\Gamma=\Sigma\Gamma_i$ is the total width of the $X(3872)$
decays into all $\{i\}$ non-$D^{*0}\bar D^0 +c.c.$ channels.
According to Refs. \cite{Belle11} and \cite{NNAEVR}  $\Gamma <
1.2$ MeV\,!

The renormalization of mass \cite{m+mX}
\begin{equation}
\label{renormalization}
 m_{\chi_{c1}(2P)}^2-m_X^2 -\Pi_X(m_X^2)=0
\end{equation}
results in
\begin{equation}
\label{Delta}
 \Delta \left (2m_X+\Delta\right
)=\Pi_X(m_X^2)\approx\left (g_A^2/8\pi^2\right
)4\ln(2\Lambda/m_+)\,.
\end{equation}

The renormalized propagator has the form \cite{exact}
\begin{equation}
\label{Rpropagator}
 D_X(s)= m_X^2-s +\Pi_X(m_X^2) -\Pi_X(s)-\imath
m_X\Gamma\,.
\end{equation}

 If $\Delta = m_{\chi_{c1}(2P)} - m_X \approx
50$ MeV, see Eq. (\ref{shiftmass})\,, then $g_A^2/8\pi\approx 0.2$
GeV$^2$ for $\Lambda=10$ GeV. According to Ref. \cite{NNAEVR} such
$g_A^2/8\pi$ results in $BR(X\to D^0\bar D^{*0}+ \bar
D^0D^{*0})\approx 0.3$ \cite{otherschannels}.

Thus, we expect that a number of unknown mainly two-gluon decays
of $X(3872)$ into non-$D^{*0}\bar D^0 + c.c.$ states are
considerable \cite{chic11P}. For details see Ref. \cite{NNAEVR}.
The discovery of these decays would be the strong (if not
decisive) confirmation of our scenario.

\begin{figure}[ht]
\begin{center}
\begin{tabular}{ccc}
\hspace*{-45pt}
\includegraphics[width=10cm,height=5.5cm]{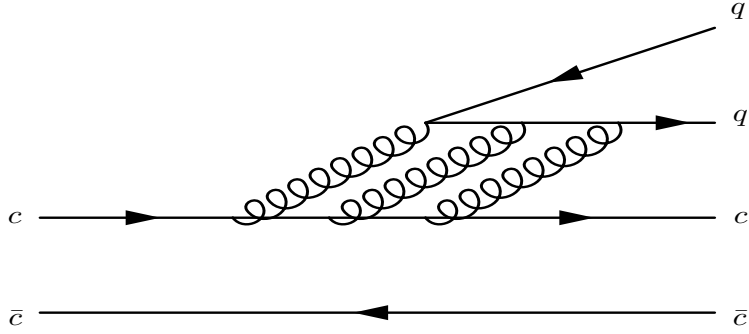}
\end{tabular}
\end{center}
\vspace*{-15pt} \caption{  The three-gluon production of the
$\omega$ and $\rho$ mesons (via the contribution $\sim m_u-m_d$ ).
 All possible permutations of gluons  are assumed. } \label{fig2}
\end{figure}

As for $BR(X\to\omega J/\psi)\sim BR(X\to\rho J/\psi)$, Eq.
(\ref{isotopicviolation}), this could be a result of dynamics. In
our scenario the $\omega J/\psi$ state is produced via the three
gluons, see Fig. \ref{fig2}.

\begin{figure}[ht]
\begin{center}
\begin{tabular}{ccc}
\hspace*{-45pt}
\includegraphics[width=10cm,height=5.5cm]{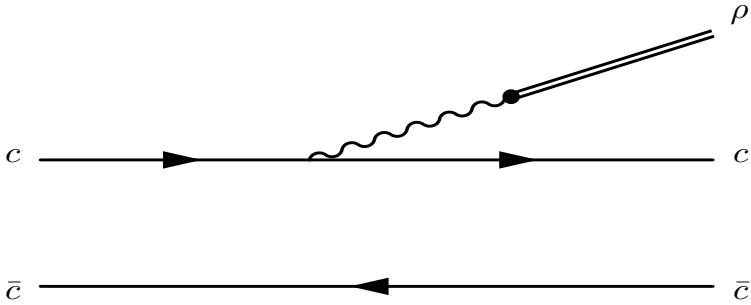}
\end{tabular}
\end{center}
\vspace*{-15pt} \caption{  The one-photon production of the $\rho$
meson. All possible permutations of  photon are assumed. }
\label{fig3}
\end{figure}

 As for the $\rho J/\psi$ state, it
is produced both via the one photon, see Fig. \ref{fig3}, and via
the three gluons (via the contribution $\sim m_u-m_d$ ), see Fig.
\ref{fig2}.
\newpage
Close to our scenario is an example of the $J/\psi\to\rho\eta'$
and $J/\psi\to\omega\eta'$  decays. According to Ref. \cite{PDG14}
\begin{equation}
\label{Jpsi}
 BR(J/\psi\to\rho\eta')=(1.05\pm 0.18)\cdot10^{-4}\ \ \mbox{and}\ \
BR(J/\psi\to\omega\eta')=(1.82\pm 0.21)\cdot10^{-4}.
\end{equation}

 Note that in the $X(3872)$ case the $\omega$ meson is produced on its
tail ($m_X-m_{J/\psi}=775$ MeV), while the $\rho$ meson is
produced on a half.

It is well known that the physics of charmonium ($c\bar c$) and
bottomonium ($b\bar b$) is similar. Let us compare the already
known features of X(3872) with the ones of $\Upsilon_{b1}(2P)$.

Recently, the LHCb Collaboration published a landmark result
\cite{LHCb14}
\begin{equation}
\label{Xtogamma}
 \frac{BR(X\to\gamma \psi(2S))}{BR(X\to\gamma
J/\psi)}=C_X\left(\frac{\omega_{\psi(2S)}}{\omega_{J/\psi}}\right)^3=2.46\pm
0.7\,,
\end{equation}
where $\omega_{\psi(2S)}$ and $\omega_{J/\psi}$ are the energies
of the photons in the $X\to\gamma \psi(2S)$ and $BR(X\to\gamma
J/\psi)$ decays, respectively.

 On the other hand, it is known \cite{PDG14} that
 \begin{equation}
 \label{chib1(2P)}
\hspace*{-18pt}\frac{BR(\chi_{b1}(2P)\to\gamma
\Upsilon(2S))}{BR(\chi_{b1}(2P)\to\gamma
\Upsilon(1S))}=C_{\chi_{b1}(2P)}\left(\frac{\omega_{\Upsilon(2S)}}{\omega_{\Upsilon(1S)}}\right)^3=2.16\pm0.28\,,
\end{equation}
where $\omega_{\Upsilon(2S)}$ and $\omega_{\Upsilon(1S)}$ are the
energies of the photons in the $\chi_{b1}(2P)\to\gamma
\Upsilon(2S)$ and $\chi_{b1}(2P)\to\gamma \Upsilon(1S)$ decays,
respectively.

Consequently,
\begin{equation}
\label{CX}
 C_X=136.78\pm38.89
\end{equation}
and
\begin{equation}
\label{Cchib1(2P)}
 C_{\chi_{b1}(2P)}=80\pm10.37\
\end{equation}
as all most lucky versions of the  potential model predict for the
quarkonia, $C_{\chi_{c1}(2P)}\gg 1$ and $C_{\chi_{b1}(2P)}\gg 1$.

 According to Ref. \cite{PDG14}
\begin{equation}
\label{omegaUpsilon(1S)}
 BR(\chi_{b1}(2P)\to\omega\Upsilon(1S))=\left
(1.63\pm^{0.4}_{0.34}\right )\%\,.
\end{equation}

If  the one photon mechanism dominates in the $X(3872)\to\rho
J/\psi$ decay, see Fig.\ref{fig3}, then one should expect
\begin{equation}
\label{onephotonrhoUpsilon(1S)}
 BR(\chi_{b1}(2P)\to\rho\Upsilon(1S))\sim(e_b/e_c)^2\cdot 1.6\,\%=(1/4)\cdot
1.6\,\%= 0.4\%\,,
\end{equation}
 where $e_c$ and
$e_b$ are the charges of the $c$ and $b$ quarks, respectively.

If  the three gluon mechanism (its part $\sim m_u-m_d$ ) dominates
in the $X(3872)\to\rho J/\psi$ decay, see Fig.\ref{fig2}, then one
should expect
\begin{equation}
\label{threegluonrhoUpsilon(1S)}
 BR(\chi_{b1}(2P)\to\rho\Upsilon(1S))\sim 1.6\%\,.
\end{equation}

We believe that  discovery of a significant number of unknown
decays of $X(3872)$ into non-$D^{*0}\bar D^0 +c.c.$ states and
discovery of the $\chi_{b1}(2P)\to\rho\Upsilon(1S)$ decay  could
decide destiny of X(3872).

Once more, we discuss the scenario where the $\chi_{c1}(2P)$
charmonium sits on the $D^{*0}\bar D^0$ threshold but not a mixing
of the giant $D^{*}\bar D$ molecule and the compact
$\chi_{c1}(2P)$ charmonium, see, for example, Refs. \cite{KR},
\cite{BAK}, and references cited therein. Note that the mixing of
such states requests the special justification. That is, it is
necessary to show that the transition of the giant  molecule into
the compact charmonium is considerable at insignificant
overlapping of their wave functions. Such a transition $\sim
\sqrt{V_{\chi_{c1}(2P)}/V_{X(3872)}}$ and a branching ratio of a
decay via such a transition $\sim V_{\chi_{c1}(2P)}/V_{X(3872)}$.

We are grateful to A.E. Bondar, M. Karliner, B.A. Kniehl, and J.L.
Rosner for useful discussions.

 This work was supported in part by RFBR, Grant No 13-02-00039,
and Interdisciplinary project No 102 of Siberian division of RAS.


\begin{thebibliography}{99}
\bibitem{m+mX} When $(m_{\bar D^0}+m_{D^{*0}})^2<s<(m_{D^+}+m_{D^{*-}})^2$\,, the renormalization of mass has the
form $ m_{\chi_{c1}(2P)}^2-m_X^2 -
 Re(\Pi_X^{\bar D^0
D^{*0}}(m_X^2)) - \Pi_X^{D^- D^{*+}}(m_X^2)=0$\,.
\bibitem{exact} The exact formulae of $Re(\Pi_X(m_X^2)) -\Pi_X(s)$ in all regions of $s$ can be
found in Ref. \cite{NNAEVR}\,.
\bibitem{otherschannels}
The assumption of the determining role of the $D^*\bar D + c.c.$
channels in the  shift of the mass  of the $\chi_{c1}(P)$ meson is
based on the following reasoning.  Let us imagine that $ D $ and $
D ^ * $ mesons are light, for example,  as  the $ K $ and $ K ^ *
$ mesons.  Then the width of $ X (3872) $ meson is equal 50 MeV
for $g_A^2/8\pi=0.2 GeV^2$  that much more than the width of its
decay into all non- $D^{*0}\bar D^0$ channels, $\Gamma < 1.2$ MeV.
That is, in our case the coupling of the $X(383)$ meson with the
$D^*\bar D + c.c.$ channels  is rather strong.
\bibitem{chic11P} Note that in the $\chi_{c1}(1P)$ case the
width of the two-gluon decays equals 0.56 MeV \cite{PDG14} that
agrees with $\Gamma<1.2$ MeV satisfactory.
\end{thebibliography}
\end{document}